\documentclass[twocolumn,showpacs,showkeys,preprintnumbers,amsmath,amssymb,superscriptaddress]{revtex4}

\usepackage{graphicx}
\usepackage{dcolumn}
\usepackage{bm}

\begin{document}


\title{Induced plasma magnetization due to magnetic monopoles}

\author{Felipe A. Asenjo}
\email{fasenjo@levlan.ciencias.uchile.cl}
\affiliation{Departamento de F\'isica, Facultad de Ciencias,
  Universidad de Chile, Casilla 653, Santiago, Chile.}
\affiliation{Departamento de Ciencias, Facultad de Artes Liberales,
  Universidad Adolfo Ib\'a\~nez, Diagonal Las Torres 2640, Pe\~nalol\'en,
  Santiago, Chile.}

\author{Pablo S. Moya}
\email{pmoya@zeth.ciencias.uchile.cl}
\affiliation{Departamento de F\'isica, Facultad de Ciencias,
  Universidad de Chile, Casilla 653, Santiago, Chile.}

\date{\today}

\begin{abstract}
When magnetic monopoles are introduced in plasma equations, the
propagation of electromagnetic waves is modified. In this work is
shown that this modification leads to the emergence of a ponderomotive
force which induces a magnetization of the plasma. As a result, a
cyclotron frequency is induced in electrons. This frequency is
proportional to the square of the magnetic charge. 

\end{abstract}



\maketitle

\section{Introduction}

A common problem in electrodynamics and plasma physics courses, which 
vividly illustrates the phenomenon of propagation of waves through a
plasma, is the calculation of the dispersion relation 
of electromagnetic transverse waves propagating in a plasma
composed by fixed ions and moving electrons. Such a problem is widely
used to introduce concepts like cutoff and plasma frequencies, as well
as, group and phase velocity of a wave.

On the other hand, it is well known that the Maxwell's equations become
symmetric when electric and magnetic charges (magnetic monopoles) are
theorized~\cite{jackson}. Even, in an elegant theoretical
demonstration, Dirac shown that the simple existence of only one
magnetic monopole in the universe leads to an explanation of the
quantization of the elementary electric charge~\cite{dirac1,dirac2}.

Once magnetic monopoles are introduced in Maxwell's equations, they
become simmetric and it is expected that new effects appear due to their
simmetry. Including magnetic charges, the equations are
\begin{equation}
\nabla\cdot {\bf E}=4 \pi \rho_e\, \quad , \quad \nabla\cdot{\bf B}=4 \pi \rho_g\, ,
\label{consMax}
\end{equation}
\begin{equation}
\nabla\times{\bf B}=\frac{1}{c}\frac{\partial {\bf E}}{\partial t}+\frac{4\pi}{c}{\bf J}_e\,
\,\, , \,\, \nabla\times{\bf E}=-\frac{1}{c}\frac{\partial {\bf B}}{\partial
  t}-\frac{4\pi}{c}{\bf J}_g\, ,
\label{eqMaxe}
\end{equation}
where $\rho_e$ is the electron charge density, $\rho_g$ is the magnetic
charge density, ${\bf J}_e$ is the electron current density, and ${\bf
  J}_g$ is the magnetic current density.

A simple test is to consider a plasma containing electric and magnetic charges. Such
analysis has great theoretical~\cite{moulin,birula} and
pedagogical~\cite{meyer} value, because of its simplicity, and the
discussion that emerges from these studies.

As a simple theoretical exercise we show how a magnetization is
induced in a plasma when magnetic charges are introduced.  
The problem is the following: we have a gas composed by magnetic
monopoles and electrons. Suddenly, an electromagnetic wave
passed through this fluid. We look for the effect in the propagation
of the very well-known simplest wave modes when magnetic monopoles are
included.

\section{Electromagnetic waves}
\label{coldw}
Let us consider a plasma as a fluid which contains electrons, with charge $-e$
and mass $m_e$, and magnetic monopoles with magnetic charge $g$ and
mass $m_g$. Let us suppose that there exist other particles with
respectives opposite electron and magnetic charges and with the same
densities, which provide the total charge neutrality of this plasma~\cite{meyer}.

The magnetic monopoles modify the well-known dispersion relation for
electromagnetic waves. In this section, we show the derivation of the relation for this
propagation mode considering the electric and magnetic charges as a
fluids and perturbing their equilibrium velocities. In order to do
this, we expand every quantity in Maxwell's equations
$\,\phi\,$ as $\,\phi_0\,+\,\phi_1\,$, where $\,\phi_0\,$ is the zeroth order (equilibrium) value
of $\phi$, and $\phi_1$ is the first order perturbation.
In that sense, throughout this work, electrons have a zeroth order
electron density  $n_{0e}$ and a first order electron density $n_e$. 
In a similar way, magnetic monopoles have an zeroth order magnetic density  $n_{0g}$
and a first order magnetic density $n_g$. Thus, the first order
electric charge density is $\rho_e=-en_e$ and the first order magnetic
charge density is $\rho=g n_g$. Thereby, the first order electric and
magnetic current density are
\begin{equation}
 {\bf J}_e=-e n_{0e} {\bf v}_e \qquad , \qquad {\bf J}_g=g n_{0g}{\bf v}_g\, ,
\end{equation}
where ${\bf v}_e$ and ${\bf v}_g$ are the first order electron and
magnetic monopole velocities respectively. Both zeroth order
velocities are null.

As in the case of a cold plasma, the classical equations of motion for
electrons and magnetic monopoles~\cite{birula}, at first order are
\begin{equation}
\frac{\partial {\bf v}_e}{\partial t}=\frac{-e}{m_e}{\bf E}_1\qquad ,\qquad \frac{\partial {\bf v}_g}{\partial t}=\frac{g}{m_g}{\bf B}_1\, .
\label{eqmotione}
\end{equation}
As seen, the Lorentz force over a magnetic monopole is the same as for an
electron but changing electric fields by magnetic field and vice versa.

In order to obtain the dispersion relation, the usual method is apply
a Fourier transform over all the linearized equations with the form
$\exp(i{\bf k}\cdot{\bf r}-i\omega t)$, where $\omega$ is the frequency and ${\bf k}$ is the wavenumber of the wave.
Thus, the equations of motion (\ref{eqmotione}) becomes
\begin{equation}
 {\bf \tilde v}_e=\frac{-ie}{m_e\omega}{\bf \tilde E}_1 \qquad ,\qquad {\bf
   \tilde v}_g=\frac{ig}{m_g\omega}{\bf \tilde B}_1\, ,
\label{eqmotionglineal}
\end{equation}
respectively. Here ${\bf \tilde v}_e$ is the Fourier transform of ${\bf v}_e$,
and the same for the other quantities.

We can write the electron and magnetic current density with the help
of the velocities given by Eqs.~(\ref{eqmotionglineal}). Therefore,
Eqs.~(\ref{eqMaxe}) can be rewritten as
\begin{equation}
 c{\bf k}\times{\bf \tilde B}_1=\left(\frac{\omega_p^2-\omega^2}{\omega}\right){\bf \tilde
   E}_1\,\, ,\,\,  c{\bf k}\times{\bf \tilde
   E}_1=\left(\frac{\omega^2-\omega_m^2}{\omega}\right){\bf \tilde B}_1\, , 
\label{eqmotionelineaju}
\end{equation}
where we define the electron plasma frecuency and a magnetic monopole plasma frecuency respectively as
\begin{equation}
  \label{eqfrequencies}
\omega_p=\sqrt{\frac{4\pi e^2 n_{0e}}{m_e}} \qquad , \qquad \omega_m=\sqrt{\frac{4\pi g^2 n_{0g}}{m_g}}\, .  
\end{equation}

Finally, the dispersion relation for a transversal electromagnetic
wave, with  ${\bf k}\cdot{\bf\tilde E}_1=0$, in an electron plasma with
magnetic monopoles can be obtained from Eqs. (\ref{eqmotionelineaju}). This gives 
\begin{equation}
c^2 k^2=\frac{1}{\omega^2}\left(\omega^2-\omega^2_p\right)\left(\omega^2-\omega^2_m\right)\, .
\label{disprelmono}
\end{equation}

Dispersion relation \eqref{disprelmono} have been obtained previously
in Ref.~\cite{meyer}. Notice that when magnetic monopoles are
neglected, $\omega_m=0$ and we recover the usual dispersion relation
$\omega^2=\omega_p^2+c^2 k^2$ for electromagnetic waves in cold plasmas.

\section{Magnetization due to ponderomotive force}

By incorporating magnetic charges can be seen several effects in the
propagation of waves through plasmas~\cite{birula,meyer}.
We focus on the emergence of a magnetic field due to the
ponderomotive force that the wave exerts on the electrons.
This magnetization appears due to the magnetic charge corrections in dispersion relation \eqref{disprelmono}. 
Similarly to the inclusion of magnetic monopoles, it has been
studied the plasma magnetization when quantum effects are considered in
the calculation of the dispersion relation of an electron plasma~\cite{jung,nshukla}.

The ponderomotive force ${\bf f}$
induced by the high-frequency electromagnetic waves in a plasma is
given in general by ${\bf f}={\bf f}^{(s)}+{\bf f}^{(t)}$,
where the ponderomotive forces ${\bf f}^{(s)}$ and ${\bf f}^{(t)}$ are
related to the space $(s)$ and time $(t)$ variations of the amplitude
$|{\bf E}_1|$ of the electric field. This force is a nonlinear effect which origin is the inhomogeneity of
the electromagnetic field.
Each force component is given by~\cite{karp}
\begin{equation}
 {\bf f}^{(s)}=\frac{\epsilon-1}{16\pi}\nabla|{\bf E}_1|^2\, ,\quad {\bf
   f}^{(t)}=\frac{{\bf k}}{16\pi\omega^2}\frac{\partial (\omega^2 (\epsilon-1))}{\partial
   \omega}\frac{\partial|{\bf E}_1|^2}{\partial t}\, ,
\end{equation}
where $\epsilon=c^2 k^2/\omega^2$ is the dielectric function of the propagation mode.
Using the dielectric function $\epsilon$ of the dispersion relation 
\eqref{disprelmono}, we can calculate the ponderomotive force for the
high-frequency electromagnetic wave. It is straightforward to obtain
an expression for the poderomotive force related to time variations
\begin{eqnarray}
 {\bf f}^{(t)}=-\frac{\omega_p^2\omega_m^2}{8\pi\omega^5}{\bf k}\frac{\partial|{\bf E}_1|^2}{\partial t}\, .
\label{pondero}
\end{eqnarray}
Notice that the effects of magnetic charges in the ponderomotive force
are in $\omega_m$. When $\omega_m=0$ the ponderomotive force ${\bf f}^{(t)}$ is null.
 
The effect of ponderomotive force of the electromagnetic
wave is to push the electrons locally. Thus, it creates a slowly
varying electric field ${\bf E}_s$, such that
${\bf f}=e n_{0e}{\bf E}_s=-e n_{0e}\nabla\phi_s-(e n_{0e}/c) \partial_t {\bf
  A}_s$~\cite{jung,nshukla}. Using the ponderomotive forces \eqref{pondero}, we can
identify the slowly varying vector potential as
\begin{eqnarray}
{\bf A}_s=\frac{c\omega_p^2\omega_m^2{\bf k}}{8\pi\omega^5 e n_{0e}}|{\bf E}_1|^2\, .
\label{vectorfield}\end{eqnarray}

Owing to the well-known relation ${\bf B}=\nabla\times{\bf A}$,
the vector field \eqref{vectorfield} induces a slowly varying
magnetic field ${\bf B}_s$
\begin{equation}
 {\bf B}_s = \frac{ec\omega_m^2}{2 m_e\omega^5}\nabla\times\left({\bf k}|{\bf E}_1|^2\right)\, .
\label{magneticf}
\end{equation}
This magnetic field interacts with electrons and induces an electron
cyclotron frequency $\Omega_{cs}=-e B_s/m_ec$. Taking the approximation
$\nabla\times({\bf k}|{\bf E}_1|^2)\approx k|{\bf E}_1|^2/L$, where $L$ is the scale
length of $|{\bf E}_1|^2$~\cite{jung,nshukla}. The induced electron
cyclotron frequency is given by
\begin{equation}
 \Omega_{cs}=-\frac{\omega_m^2 k}{2\omega^3 L}V_0^2\, ,
\label{eleccyclomm}
\end{equation}
where $V_0= e|{\bf E}_1|/m_e\omega$ is the electron quiver velocity. 
Notice that when magnetic monopoles are neglected, $\Omega_{cs}=0$ and there are no magnetization.

\section{Conclusions}

Starting from the inclusion of magnetic charges in Maxwell's
equations, we have studied the simple problem of propagation of
electromagnetic waves in a cold electron plasma. We have shown that
when magnetic monopoles are introduced in an electron plasma, the
dispersion relation~\eqref{disprelmono}, and thereby the propagation
of electromagnetic waves, is modified because of the magnetic monopole plasma frequency~\eqref{eqfrequencies}.

Due to the presence of the magnetic monopoles, the plasma is
magnetized with a magnetic field given by Eq.~\eqref{magneticf}, which is of $g^2$ order. However, it can
interact with the electrons inducing an electron cyclotron frequency.
The frequency  \eqref{eleccyclomm} is only due to the ponderomotive force related to time variation of the field intensity and which is induced by
magnetic charges. It depends on the magnetic monopole plasma frequency $\omega_m$ and on the propagation mode 
\eqref{disprelmono} through $\omega$. Finally, notice that the magnetization and the cyclotron motion decrease when the frequency of the electromagnetic wave increase. 

The above calculations of the dispersion relation and of the plasma magnetization are very simple. Therefore, they are useful as a pedagogical tool for teaching the recognition of basic phenomena arising of magnetic charges in plasmas, and of how these simple effects can bring new insights of the electron plasma dynamics at classical level.

\begin{acknowledgments}
P. S. M. thanks to CONICyT, Chile Doctoral Fellowship for their financial support.
\end{acknowledgments}

\end{document}